# Customizing an Affective Tutoring System Based on Facial Expression and Head Pose Estimation

Mahdi Pourmirzaei, Gholam Ali Montazer, Ebrahim Mousavi

{m.poormirzaie, montazer, e.moosavi }@modares.ac.ir

Tarbiat Modares University

## Abstract

In recent years, the main problem in e-learning has shifted from analyzing content to personalization of learning environment by Intelligence Tutoring Systems (ITSs). Therefore, by designing personalized teaching models, learners are able to have a successful and satisfying experience in achieving their learning goals. Affective Tutoring Systems (ATSs) are some kinds of ITS that can recognize and respond to affective states of learners. In this study, we designed, implemented, and evaluated a system to personalize the learning environment based on the facial emotions recognition, head pose estimation, and cognitive style of learners. First, a unit called Intelligent Analyzer (AI) created which was responsible for recognizing facial expression and head angles of learners. Next, the ATS was built which mainly made of two units: ITS, IA. Results indicated that with the ATS, participants needed less efforts to pass the tests. In other words, we observed when the IA unit was activated, learners could pass the final tests in fewer attempts than those for whom the IA unit was deactivated. Additionally, they showed an improvement in terms of the mean passing score and academic satisfaction.

## 1. Introduction

An Intelligent Tutoring System (ITS) is a hybrid system that uses behavioral, cognitive, psychological, educational, and computer sciences to enhance learning in an e-learning environment. The term "intelligent" in ITS refers to the ability to adapt the learning environment to the learner's personal characteristics such as learning speed and domains where the learner exhibits superior or weaker learning ability. Researchers believe that the performance of an ITS can be significantly improved by considering the learner's affective state in the learning environment [1], [2]. Affect refers to the neurophysiological state of experiencing a particular feeling such as emotion. Thus, emotion is an affective state associated with our neurophysiological response to stimuli [3]. One of the important determinants of the success of personal (private) education is the teacher's ability to detect and respond to the affective state of the learner [4]. In a study [5], it has been shown that only a few positive changes in mental state can make people more creative, flexible, precise, and efficient in solving problems. Despite this evidence, the majority of existing ITSs ignore the affective state of the learner. This can be due to the complexity of the concept of affect and the difficulty of detecting affective states in educational settings [6]. Systems that take into account the learner's affective state in their educational strategies are called Affective Tutoring Systems (ATSs). These systems have been developed to mimic the behaviors of a real teacher in adapting the learning environment to the learner's affective state [7]. There are two main challenges in developing an ATS [8]: (1) how to detect the affective state of the learner, and (2) how to adapt the learning environment to the learner's affects.

The most important indicators of people's emotions and affective states are their facial expressions. One of the most popular methods of emotion recognition from facial expressions is the Ekman model [9]. In this model, emotions are divided into eight categories: enjoyment, sadness, contempt, surprise, fear, anger, disgust, and neutral. Despite its popularity, this method does not work well in learning because it cannot identify the intensity of positive and negative emotions. Also, important emotions associated with fatigue, confusion, and anxiety cannot be detected and measured by the Ekman method. Furthermore, emotions like fear, sadness, and disgust are rarely experienced in the learning environment [10]–[12].

Considering the above issues, this study uses Russell's two-dimensional model for emotion recognition [13]. This model defines emotions in two dimensions: Arousal (Activation vs. Deactivation) and Valence (Pleasant vs. Unpleasant). In the diagram of this model, the vertical and horizontal axes represent arousal and valence spectra, respectively, with the center representing the neutral state (Figure 1). Using the Russell model, a person's emotion at any level can be represented on the arousal and valence spectra. The reason for using this model rather than the Ekman model is its higher effectiveness and accuracy in educational settings, where the recognition of certain positive and negative emotions and their intensity has shown to be more important [11], [14].

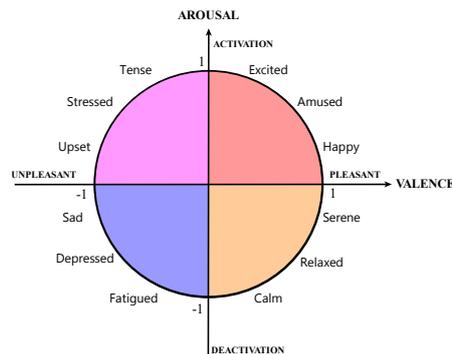

Figure 1. Diagram of Russell's two-dimensional emotion model [13].

In this study, the learning environment was adapted based on the learner's cognitive style, emotion, and head pose while watching educational content. The learner's cognitive style was identified using the method of a work [15], which assesses the wholistic-analytical and verbalizer-imager dimensions. Here, only the analytical and wholistic dimensions of the cognitive style were considered. Thus, people were divided into three categories of analytic, wholistic, and middle, and the curriculum was personalized for each category. Also, a combination of visual, aural, and written content was used for all learners.

The learner's state was determined by fusing the prediction of three methods (face recognition, emotion recognition, and head pose detection) with the help of webcams while the learner consumed the educational content. The results were compiled into a system that could take a large number of videos of the learner watching a lesson and place the person in one of 21 learning states. This state affected two things: (1) the textual feedback that will be displayed to the learner at the end of each lesson; (2) the supplementary content that will be recommended to the learner if needed.

This study was carried out in two phases:
1. Design and implementation. This phase involved designing the ITS, designing the intelligent analyzer (IA) unit (the part of the system that modifies the learning process according to the emotion and head pose), and finally implementing the designed ATS.
2. Evaluation. In this phase, the designed ATS was evaluated by preparing educational content suitable for each cognitive style and presenting them to a number of learners with and without the IA. In this step, performance was measured in terms of "academic success" and "academic satisfaction".

## 2. Related work

As mentioned in the introduction, it could be very difficult to design an effective ATS, as this will require combining psychology with educational and computer sciences [16]. One of the most important challenges in this area is how to detect affective states.

In many ATSs, affective states are primarily detected through emotion recognition from facial expressions [8], [9]. In a study [17], this approach was used to create an ATS for a Massive Open Online Courses (MOOCs). In this ATS, the learner's affective state while studying the content was determined through facial expression recognition based on the Ekman method and the adaptation was done based on predetermined rules according to the identified state. In the system proposed in a work [18], in addition to emotion recognition from the face, a module called "Semantic Clues Emotion Voting" was used to detect the learner's affective state based on a glossary of keywords that students use in the classroom, and ultimately the content was modified based on a combination of the results of the two methods.

A person's affective state can be estimated not only from the face but also through the method called sentiment analysis [19], [20] and the analysis of biometric information [21]. In a study [20], comments and feedbacks of a group of learners in an exam were analyzed and classified into two categories of positive and negative sentiments and the results were used to evaluate the quality of exercises and design better exams. In the system proposed in a study [19], the learning process continues if the learner's emotion was assessed to be positive. Otherwise, the system started asking questions about positive emotions to encourage the learner to think positively. In another ATS [21], the learner's affective state was measured by biometric technologies and the obtained information was used to tailor the educational content to each user.

## 3. Design and implementation of ATS

This phase of the work comprised three steps: (1) designing the ITS; (2) designing the IA unit; (3) implementing the ATS.

In the following subsections, we explain how the learning environment is designed, including how the learning is modified for each cognitive style, how aggregator and analyzer components are added to the system in the form of an IA, and finally how the designed ATS is implemented. The main components of the proposed system are illustrated in Figure 2.

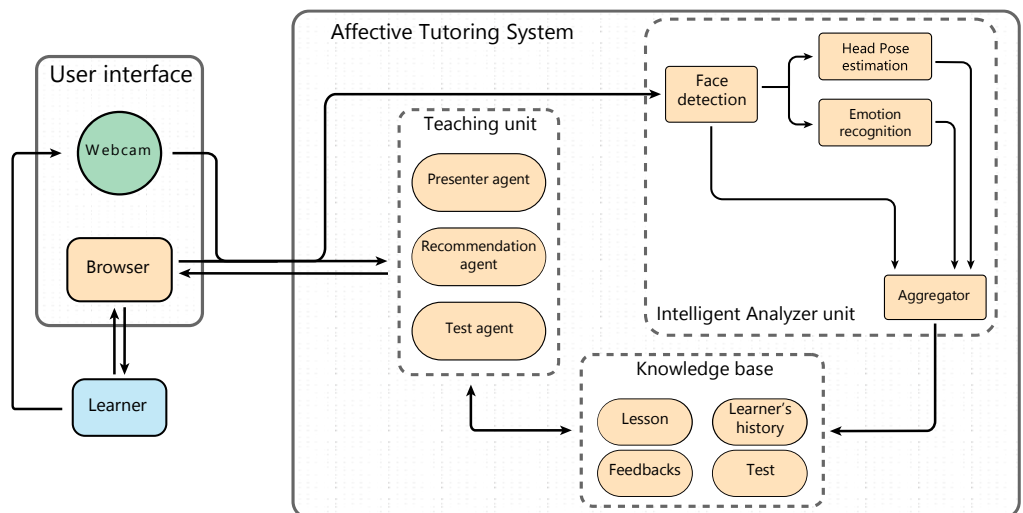

Figure 2. The parts of the proposed web-based learning system. The ATS part is on the server-side and the user interface part is on the client-side. ATS consists of three units: intelligent analyzer, teaching unit, and knowledge base.

### 3.1. ITS

The design of this unit consists of two parts: (1) design of the educational environment, and (2) personalization of the environment based on the learner's cognitive style.

The e-learning process in this learning environment involves watching and study of multiple sessions, each comprising a number of lessons. The goal is to personalize this environment for each learner. This personalization is done in two ways, which are explained below.

One of the most common ways of personalizing educational content is to consider the learning styles. While researchers have proposed a variety of different learning style models -assuming that people are different in terms of the ability to learn through visual, aural, read/write, and kinesthetic sensory modalities (VARK) -, there are still many doubts about the effectiveness of the methods that prioritize individual learning styles. In fact, despite the popularity of this subject, several studies have shown that there is little relationship between learning styles and academic achievement [22]–[27]. Meanwhile, it has been shown that using visual, aural, and read/write modalities at the same time is more likely to lead to successful learning [22], [28].

In a study [15], two dimensions were defined for the learner's cognitive style: (1) the wholistic-analytical dimension, which indicates whether a person tends to process information in whole or in parts; (2) the verbalizer-imager dimension, which indicates whether a person tends to process information in verbal form or in visual form and can be considered a subset of VARK. Considering the shortcomings of VARK learning styles, in this study, we only use the first dimension (i.e. wholistic-analytical) to classify learners, and use a combination of visual, aural, and written contents for all of them. In other words, for every cognitive group, content related to each lesson is provided in the form of a video containing text, video, and audio.

In the first personalization method, educational content is tailored to three cognitive styles, The learner's cognitive style is detected using the test provided in a study [29], which divides people into two categories of wholistic and analytical. People whose test results do not put them squarely in either of these categories (wholistic or analytical) are placed in a third category called "middle". Accordingly, learners will be classified into the three following groups:
1. Analytical
2. Wholistic
3. Middle

Three groups of lessons, one for each cognitive style, are created. For learners in the middle group, educational content is a mixture of the contents prepared for wholistic and analytical learners. More detailed information about the content is provided in Section 4-1. Each learner can only access the content prepared specifically for his or her cognitive style category. The curriculum is designed such that different groups might not have the same number of lessons at each session. Figure 3 shows an example of how personalized content is created.

The second personalization method is the analysis of the behavior of learners in videos. This is done by a unit named IA, which will be described below.

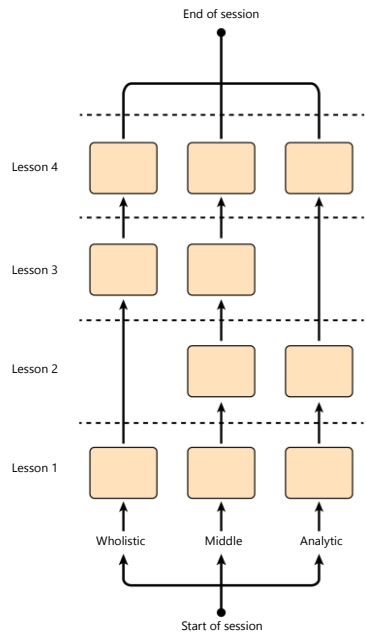

Figure 3. Lesson arrangement for learners in wholistic, analytical, and middle cognitive groups. In addition to the arrangement of the sessions, the content of the lessons might vary for each group.

### 3.2. Intelligent Analyzer

The purpose of the IA designed in this study is to monitor the state of learners while they consume the content and then, give the system the information it needs to personalize the content and feedback for each learner (Figure 4). This unit consists of two components, analyzer and aggregator. The analyzer is first designed for the analysis of each frame and then, extended for analyzing videos. The final analyzer - video analyzer - is combined with an aggregator component to create the IA. In the following, first, the methods of face detection, head pose estimation, and emotion recognition are explained, and then, creating of the analyzer and aggregator components are described.

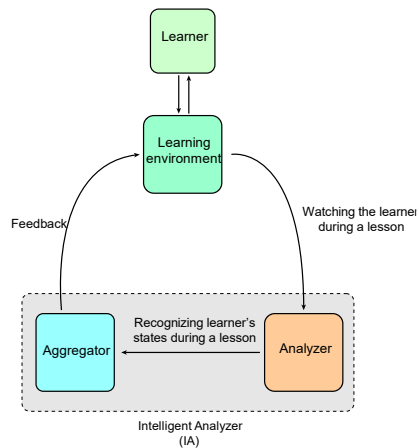

Figure 4. Relationship of the intelligent analyzer with the learning environment. The intelligent analyzer unit is the product of the fusion of the analyzer (intelligent methods) and aggregator components.

**Face Detection**: To monitor the emotion and head pose of learners, it is necessary to detect the exact location of their faces in each frame. In ATSs, face detection must be done instantly for multiple learners at the same time. Therefore, the face detection method should be able to process frames instantaneously with acceptable precision. In this study, the FaceBoxes model [30] is used for this purpose. Comprised of multiple layers of neural networks, this model is designed to perform high-speed face detection processing on GPU and CPU. This model takes an image as input and gives the exact location of faces in the frame as output.

**Head Pose Estimation**: In addition to the learner's affective state, his or her head pose relative to the webcam can also provide valuable information about the person's learning condition. For example, the head pose can help determine whether the learner is focused on the content (if the head is oriented significantly away from the webcam for a prolonged period, it might indicate a lack of focus). In this study, in addition to the affective state, the head pose is also considered in the learning personalization. To determine the head pose, three Euler angles need to be measured: yaw, pitch, and roll. Each of these angles is measured with respect to the coordinate of the origin, which here is the webcam lens. The yaw, pitch, and roll angles measured by the face detection component are indicated in Figure 5.

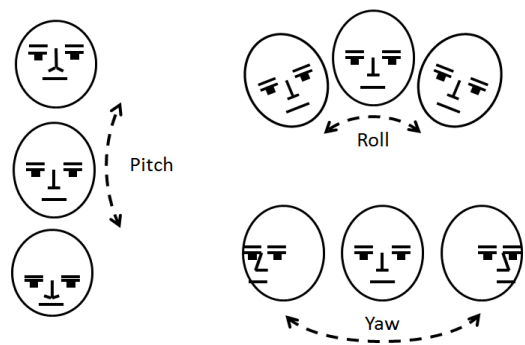

Figure 5. The outputs of the Head Pose estimation network: roll, pitch, and yaw.

Following the approach used in a work [31], in this study, the head pose is estimated with a network method based on the HopeNet [32], which takes cropped images of the face as input. In the HopeNet architecture, the exact yaw, pitch, and roll values are converted to classification labels and then the regression values are obtained from the mathematical expectation of classification heads. Here, self-supervised learning with auxiliary heads is also performed to achieve a lower average error. As mentioned in the main work [31], the self-supervised method is of the puzzling type besides three supervised head pose heads. All training settings and datasets have been derived from [31] and the deep model used for training is ResNet50 [33]. This network takes the cropped image of the head as input and gives its yaw, pitch, and roll angles as outputs.

**Emotion Recognition**: As mentioned, emotion recognition is performed using Russell's two-dimensional model. For this purpose, the method of a work [34] is used to design and train a neural network for recognizing the valence and arousal values. Like the head pose estimation network, this method is based on HopeNet. It takes face images as input, and in addition to detecting the two axes of emotion as output, uses self-supervised learning with auxiliary heads to reduce the average error. In this method, Efficientnet-B0 architecture [35] and AffectNet images [36] are used for training. This network operates by taking the cropped images of the face as input and returning two values for valence and arousal as outputs.

### 3.2.1. Analyzer design

The overall process of the IA is shown in Figure 5. This unit consists of two components, analyzer and aggregator. The analyzer is created by combining face detection, emotion recognition, and head pose estimation networks. In the following, we first explain how the process and the methods are applied to one frame (Figure 6) and then extend the process to the processing of a video (Figure 7).

Upon receiving a frame as input, the face detection algorithm can respond in three ways: (1) not detecting any faces, which will cause the processing procedure to stop; (2) detecting multiple faces, which will also cause the processing to stop, as personalization is supposed to be done for each person individually; (3) detecting a single face, in which case, the face will be cut out of the original image and converted to the appropriate size for use in the next two networks.

In the next step, the cropped image is given to the head pose estimation network, which analyzes the head pose and estimates its pitch and yaw angles.

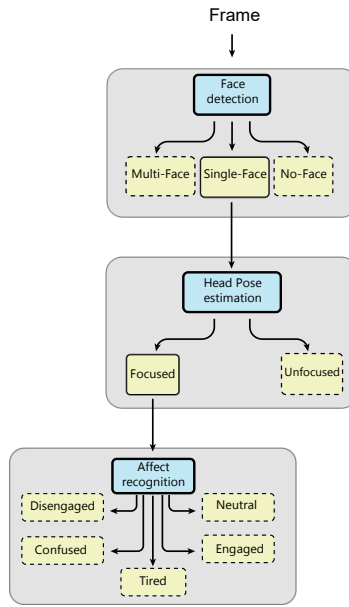

Figure 6. Flowchart of frame processing in the analyzer. At the end each frame will belong to one of eight states marked by dotted lines.

In the next step, the obtained angles are compared with a set of thresholds. If the angles fall within the defined ranges, it will be interpreted as the learner being focused on the content, but if they do not, it will mean that the learner is not focused and the processing will stop. The thresholds used for this focus range are provided in Table 1. If the angles suggest that the learner is focused, the image will be given to the emotion recognition network. This network produces two outputs, arousal value and valence value, each ranging from -10 to 10. These values are used to place the person's emotional state into one of five classes of engaged, tired, confused, disengaged, and neutral based on the two-dimensional circumplex model (Figure 7). This classification is done based on the thresholds $\alpha_1$ to $\alpha_6$, the values of which have been obtained through experimental research on different genders, ethnicities, and appearances. The values of these parameters are provided in Table 1.

Table 1. Threshold values in the intelligent analyzer.

| Threshold | value |
|---|---|
| Face Detection confidence | 0.7 |
| No-Face / Total | 0.25 |
| Multiple-Faces / Total | 0.25 |
| Unfocused / Total | 0.35 |
| Focus Yaw Range | -29 to 29 |
| Focus Pitch Range | -37 to 16 |
| $\alpha_1$ | -1.5 |
| $\alpha_2$ | 1 |
| $\alpha_3$ | 1 |
| $\alpha_4$ | -2 |
| $\alpha_5$ | 6 |
| $\alpha_6$ | -5 |
| Emotion Multiplier | 1.0 |
| Disengaged# - Aggregator | 0 |
| Engaged# - Aggregator | 1 |
| Tired# - Aggregator | 2 |
| Confused# - Aggregator | 2 |
| Multiple-Faces# - Aggregator | 2 |
| No-Face# - Aggregator | 2 |
| Numerous No-Faces# - Aggregator | 4 |
| Unfocused# - Aggregator | 2 |

While each frame can be processed on its own, the goal is to use the IA in ATS to detect the behavior of learners in videos. Therefore, for the system to work as intended on the videos, it is necessary to slightly change the processing procedure and the way states are detected by the networks.

The procedure of video processing in the analyzer is shown in Figure 7.

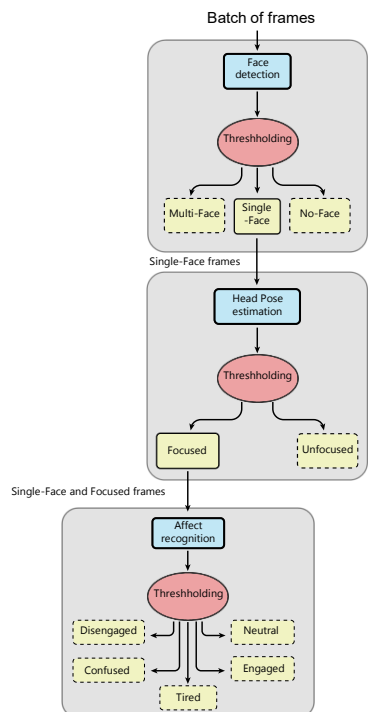

Figure 7. Flowchart of video processing in the analyzer.

Each video is converted to 15 frames per second one by the way. Assuming that the video is 10 seconds long, this conversion provides 150 frames for each video. In the first step, all frames of one video are given to the face detection network, where the output for each frame can be no face, one face, or multiple faces like the previous method. If the ratio of the number of No-Face outputs to all frames for that video is greater than the threshold given in Table 1, the video will be labeled "No-Face" and no further processing will be performed. Otherwise, if the ratio of the number of Multiple-Faces outputs to all frames for the video is greater than the threshold given in Table 1, the video will be labeled "Multiple-Faces" and if not, the video will be labeled "Single-Face" and the processing proceeds to the next step. Next, the square enveloping the face is cut out of the image and is given to the head post estimation network. To be precise, the frames containing a single face are labeled according to

Table 1. If the ratio of the number of "Unfocused" labels to the number of all labels is greater than the threshold given in Table 1, the state will be recognized as "Unfocused" and if not, the processing proceeds to the next step. The frames labeled "Focused" are given to the emotion recognition network, where the arousal and valance values are going to be calculated for each frame. All resulting valance and arousal values will then be averaged to obtain two mean valance and arousal values. Figure 8 indicates the circumplex diagram for determining the emotional state based on these two values. As can be seen, the person's emotional state can be recognized to be one of the following: engaged, tired, confused, disengaged, and neutral.

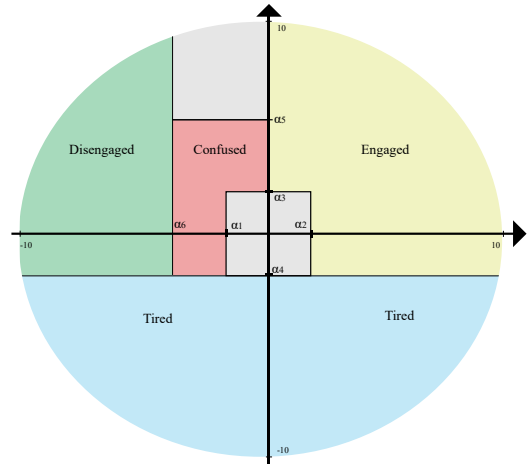

Figure 8. Circumplex diagram for determining the learners' emotional state based on arousal and valance values. The parameters are obtained from the analysis of different faces during watching educational content. Gray areas are neutral zones.

Note that in Table 1, the thresholds with the term aggregator are related to the aggregation stage and the rest are related to the video analyzer stage.

### 3.2.2. Aggregator design

While the analyzer can now take multi-second videos and determine which of the eight states they belong to, each lesson tends to last at least several minutes. A multi-minute video cannot be given to the analyzer at once, due to the fact that it requires a lot of processing and might also contain a shift in the learner's affective state, in which case the system will miss one of the states exhibited in the video. To solve this problem, the video recording process is considered to be in the form of 10-second sampling. Thus, we will ultimately have for each lesson as many states as there are 10-second video samples from that lesson. Each 10-second video sample can be labeled as one of the eight mentioned states by the way. Next, the multiple states obtained for each lesson will be combined in the aggregator. The aggregator selects the aggregation of the states based on how many 10-second videos indicate each state and whether this number exceeds the corresponding threshold in Table 1. For example, the final state "Tired+Confused" will be selected if more than two 10-second videos indicate "Tired" and "Confused".

The aggregator checks the final state in the following order:
No-Face + Multiple-Faces; Multiple-Faces; Numerous No-Faces; Tired + Unfocused; Tired + Confused; Unfocused + Confused; Engaged + Tired; Engaged + Confused; Disengaged + Confused; Tired + No-Face; Tired + Disengaged; Engaged + No-Face; Disengaged + No-Face; Engaged + Unfocused; No-Face; Unfocused; Tired; Engaged; Confused; Disengaged; Neutral.

The important point in the selection of the final state out of the above 21 options is to check them in the above order from top to bottom (Figure 9). This order is determined by trial and error, after removing triple and higher combinations as well as low probability states. More detailed information about the states is provided in the appendix. Figure 9 shows the process of state selection in the aggregator.

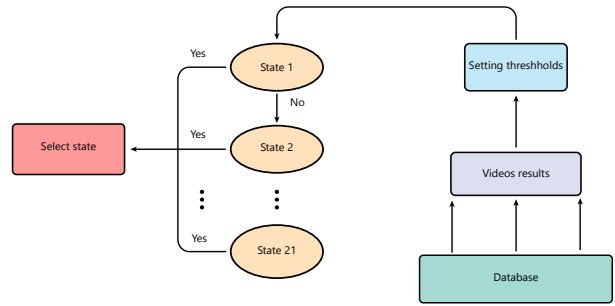

Figure 9. Flowchart of final state selection in the aggregator.

Ultimately, the IA is designed to recognize the state of learners in every lessons. Inside the IA a bunch of 10-second (or shorter) videos are processed by analyzer and then all results are combined using the aggregator. The exact operational procedure of the IA in the learning environment is shown in Figure 10. As the figure illustrates, first, the videos are processed separately and their results are stored in the database. Once a lesson is complete, the aggregator retrieves the results of all the videos recorded by the webcam from the database and aggregates them. In the end, the final state is reported back to the learning environment.

### 3.3. Implementation of ATS

To implement the ATS, the IA is embedded in the ITS. The resulting system consists of three main units (Figure 2):

**Intelligent Analyzer**: the purpose of this part is to continuously monitor the state of each learner while they watch the lessons. This component has two types of outputs: (1) a feedback message that is displayed to the user, (2) a recommendation of supplementary content for each lesson. Since the number of lessons varies in each cognitive style group, so does the supplementary content that can be recommended for each lesson. If the output of the IA is Unfocused + Confused, Engaged + Confused, Disengaged + Confused, or Confused, this unit notifies the teaching unit that the information of the supplementary content for the lesson (if there is any) must be displayed to the learner along with a message.

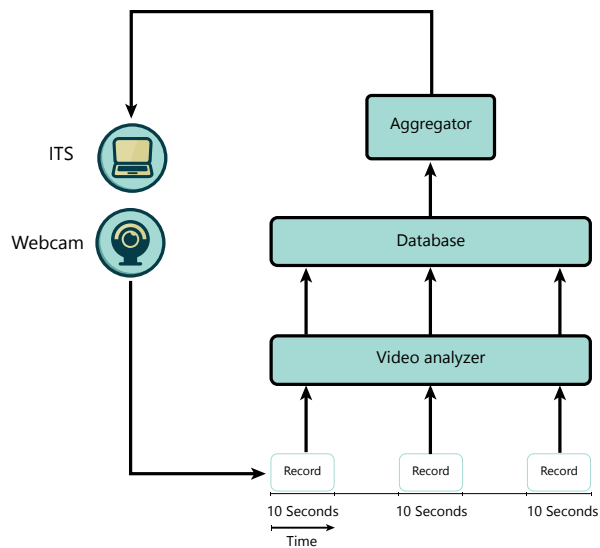

Figure 10. Work flowchart of the intelligent analyzer unit in the learning environment.

**Knowledge Base**: this is a database comprised of multiple sections. The records section contains the records of every learner including their cognitive styles, the history of lessons watched and related feedback, and the history of test results. The lesson section contains the educational content for each cognitive style. The test section contains the tests of each session, and the recommendations section contains all the feedback and recommendations that can be provided depending on the state identified by the IA. Out of the 21 states that can be detected by the IA can, three states elicit the recommendation of supplementary content or textual feedback if no supplementary content is available. The message displayed for all recommendations is in the singular first person in order to create a more person-to-person teaching experience. The text of the recommendation messages is provided in the appendix.

**Teaching Unit**: In this unit, the presenter agent retrieves the person's cognitive style from the records section (determined before login) and prepares the lessons for the person accordingly. Once the lesson is watched and the person's state is identified by the IA, the recommendation agent prepares the appropriate recommendation for display. It also prepares the link to the supplementary content when needed. The test agent prepares the test part of the session when all the lessons are over. If the person does not earn a passing grade, this agent will display all the supplementary contents of that session on the session page.

The user interface component is designed outside the learning system. It allows people to view the content of each course in their browser and interact with ATS using mouse, keyboard, and webcam. This component uses the webcam to take videos of the learner's face while they watch the content and send them to the learning system.

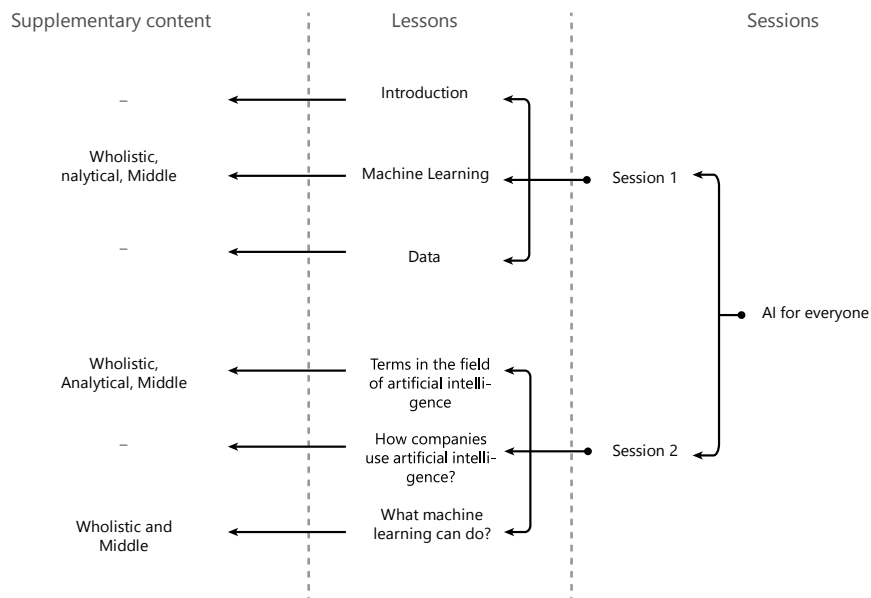

Figure 11. Course information.

## 4. Evaluation of ATS

To evaluate the system, we designed a course with contents adapted to three cognitive styles. This course was presented to a number of participants, who were divided into three groups based on their cognitive styles, and in the end, the performance of the system in terms of academic success and academic satisfaction was evaluated. These steps are described below in more detail.

### 4.1. Educational content

To evaluate the learning system, a course named "Artificial Intelligence for everyone" was prepared. In addition to evaluating the learning system, the purpose of this course was to provide an introduction to the concepts of artificial intelligence with a completely non-specialized approach. Figure 11 displays the arrangement of sessions in this course. As explained in Section 3-1, the test [29] was used to put each learner in one of three categories of analytical, wholistic, or middle before they log into the system.

In wholistic group, the educational content was created such that they gained a general picture of what they were supposed to learn in the session before proceeding to specific session. Also, the content for these learners was modified to contain less detailed information. However, to compensate the lack of detail in the lessens, supplementary contents prepared for these learners were more extensive and contained more examples and details than those prepared for the analytical group.

In analytical group, the educational content was created such that they went through the topics methodically step by step and received an overview of the covered topics at the end of each lesson. Furthermore, they received an overview of the lessons at the end of each session as well. In other words, contrary to wholistic learners, analytical learners got a general picture of what they have supposed to learn at the end of the sessions. Supplementary contents prepared for this group were not as extensive as those for the wholistic one.

For learners placed in the middle group, the contents prepared for wholistic learners were modified to contain slightly more details. Supplementary contents prepared for this style were similar to those for wholistic learners. The supplementary content prepared for each video was the same as the main content, except that the concepts were summarized and explained slower. These supplementary contents were created by providing simpler explanations with multiple examples for each lesson. As explained earlier, it is important for the supplementary contents to be hidden by default. In other words, they should become accessible under two conditions: (1) If the system recognizes that the main content is confusing or ambiguous for the learner; (2) If the learner does not earn the minimum score in the final exam of the session. In both cases, the supplementary content will appear on the lessons page.

### 4.2. Learners

Participants were chosen from a group of volunteering undergraduate and graduate students, who had little to no knowledge about artificial intelligence and were completely unfamiliar with the topics to be discussed in the course. The characteristics of the participants and the results of their cognitive style test are shown in Table 2. All participants were selected through social networks by random selection from 30 volunteers. The participants were Persian speakers residing inside and outside Iran.

Table 2. Characteristics of participants in the course.

| Total Number | | 14 |
|---|---|---|
| Education level | Bachelor's | 8 |
| | Master's | 6 |
| Age | Minimum | 19 |
| | Maximum | 28 |
| | Mean | 23.6 |
| Gender | Male | 6 |
| | Female | 8 |
| Cognitive style | Wholistic | 9 |
| | Analytical | 4 |
| | Middle | 1 |

### 4.3. Evaluation

The performance of the educational system and environment was evaluated in terms of "academic success" and "academic satisfaction" [8], which refer to the effectiveness of ATS on the participants' performance in the learning process and their satisfaction with the learning environment respectively. In order to examine these two factors, two random persons from each cognitive group were chosen as the "control group". These learners could only watch the videos and did not have access to the feedbacks of the IA unit. The rest of the participants, for whom the IA was activated, were named the "test group". Academic success was measured with the following criteria: the test score of each session, the average score in the retaken tests, and the average number of attempts to pass the test of each session. Academic satisfaction was measured by a questionnaire that was administered at the end of the course. The test results for the first and second sessions are shown in Tables 3 and 4.

Table 3. Results of the participants in the first session.

|  | Mean time spent watching the contents (minutes) | Mean number of attempts to earn a passing score | Mean passing score (%) | Mean score in the first attempt (%) | Mean score in the second attempt (%) |
|---|---|---|---|---|---|
| Control group | 71 | 2.9 | 89.9 | 72.2 | 83 |
| Test group | 112 | 2.4 | 83.7 | 65 | 81.3 |

Table 4. Results of the participants in the second session.

|  | Mean time spent watching the contents (minutes) | Mean number of attempts to earn a passing score | Mean passing score (%) | Mean score in the first attempt (%) | Mean score in the second attempt (%) |
|---|---|---|---|---|---|
| Control group | 68 | 3.2 | 86.6 | 67.1 | 79.8 |
| Test group | 98 | 1.9 | 94.4 | 81 | 100 |

At the end of the course, the participants were given a questionnaire containing three questions. The first and second questions were about the quality of educational content and how easy or hard it was to work with the system (Table 5). In the third question, they were asked to describe their experience of working with the e-learning system and the problems they encountered during the course.

Table 5. Results of the administered questionnaire.

|  | Mean for responses given the first question | Mean for responses given to the second question |
|---|---|---|
| Control group | 8.2 | 7.9 |
| Test group | 7.7 | 8.7 |

The results of the evaluation of the system in terms of academic satisfaction and academic achievement were as follows:

- People in the control group (for whom the IA was deactivated) needed more attempts to pass the tests in both sessions. For the first session, the average number of attempts in the control group was 11% higher than in the test group. For the second session, this figure was 68%.
- According to the responses given to the third question of the academic satisfaction questionnaire, in the first session, people in the test group had a sense of excitement for receiving feedback from the system and spent a lot of time testing the feedbacks instead of focusing on the educational content. This might explain the significant difference between the two groups in terms of the mean time spent watching videos and the mean test score in the first session. In the second session, however, people in the test group, who were then familiar with the feedback mechanism, paid more attention to the learning process. As a result, they showed significant improvement in terms of the mean passing score, the mean number of attempts to earn a passing score, and the mean time spent watching videos.
- While test and control groups were given the same content, people in the test group had a more favorable view of the quality of the content. This shows the effect of the intelligent learning system on the participants' perception of and interest in the provided content. Although the IA was disabled for the control group, two people in this group stated that the learning environment was intelligent and was able to present the lessons in a comprehensible way. This perception could be attributed to the provision of supplementary contents upon failure to earn the passing score and the personalization of content based on the learner's cognitive style.

## 5. Conclusion

In this study, considering the learner's affective state, head pose estimation, and cognitive style had an impact on the learning process, an ATS was designed, implemented, and evaluated. In other words, the proposed system personalized the learning environment based on facial emotions, head pose, and cognitive style of learners. The results of performance evaluations showed when the Intelligent Analyzer unit - a unit that responsible for recognizing learner's state based on the face detection, head pose estimation and facial emotion recognition networks - was activated, learners could pass the final tests in fewer attempts and had 10% higher academic satisfaction than those for whom the Intelligent Analyzer unit was deactivated.

# Appendix

## A. Implementation details

The designed ATS has been built in the form of a website that can be accessed via the Internet. The website operates as follows. First, the "admin" account defines the user accounts for the system. Prior to logging into the education environment, learners will be asked to partake in a cognitive style recognition test, which determines their cognitive style group. Based on the test results, each user will fall into one of three categories: wholistic, analytical, or middle, which will be specified in the accounts. The purpose of this categorization is to personalize the content offered to each learner based on their cognitive style. The tutoring system operates according to the following rules:

- To start a course, at least one session needs to be created;
- Once sessions are created, three cognitive style groups will be automatically created for each session. The video content specific to each cognitive style group then needs to be provided. In other words, while the educational contents of all groups in each session will be created all at once, different videos and tests can be defined for different groups;
- Only the admin account can create sessions and define videos and tests;
- To be allowed into the system, the user must turn on the webcam. Otherwise, access to the content will be blocked;
- Users cannot access the contents and tests of other cognitive style groups;
- After each video, the results obtained from the processing of webcam images will be aggregated and the user will be given feedback accordingly. This feedback can be a recommendation to watch supplementary videos;
- Each video of each cognitive style group can have one or more supplementary videos;
- Supplementary videos are hidden by default and will appear as a link alongside the feedback on the lessons page once the recommendation conditions are met;
- Each session consists of several videos (lessons) ending with a four-option multi-choice test;
- If multiple sessions have been created, each session can only be accessed upon earning a passing score in the test of the previous session;
- Upon failing a test, the user will be shown the supplementary videos recommended for that session on the lesson page;
- Users can take each test multiple times to earn a passing score;
- The admin account can view a graphical representation of the results of the intelligent analyzer for users;
- The intelligent feedbacks the system gives to users are just recommendations. Thus, users can ignore them and proceed to the next video;
- Users can view their image by pressing the "display webcam" button.

To start working with the tutoring system, users must enter their account information on the login page in order to access the session page of their account. Once logged in, users must select a session to access the corresponding video page and then watch all the videos to the end.

While the user is watching the videos, the system takes video records of his or her face via the webcam based on a schedule. During lessons, the "display webcam" button will be disabled by default to avoid distracting users. In the default video recording schedule, the system records 10-second videos with 10-second pauses between them (this schedule is adjustable and must be set based on the bandwidth and processing speed. For the system implemented in this study, taking 10-second records with 10-second pauses was determined to be the best choice). The video records will be converted into a format that is suitable for processing and then will be sent to the server for analysis by the intelligent analyzer. At the end of each educational video, all records will be processed, saved, and aggregated. After some moments, the intelligent system will show a message containing a recommendation for the user at the top of the video window.

After watching all the videos of a session, the user must take the final test of the session. The passing score is 80 out of 100. Upon earning the passing score, the user will be given access to the next session. For users who fail to earn a passing score, the supplementary videos of that session will appear in the recommendation massage to help them improve their scores.

## B. Intelligent analyzer

Table 1 shows the processing time of different components of the intelligent analyzer. As this table shows, it takes Nvidia-GTX 1080Ti graphics card about 4 seconds to process a 10-second video record of the system. Since videos are recorded with 10-second pauses, four users can be analyzed simultaneously in real-time.

Table 1 Processing time of the components of the intelligent analyzer for video records

|  | Time (ms) |
| --- | --- |
| Face detection | 1.8 |
| Head Pose estimation | 6 |
| Emotion recognition | 6 |
| Pre-processing | 10 |
| Total | 23.8 |

## C. Aggregator

The state of the learner and the text message that must be displayed for each state are determined in the aggregator unit of the system. For the states involving confusion, the system has two sets of messages, one for when the lesson has supplementary content, and another for when it has not:

**No-face+Multiple-Faces**: To receive accurate feedback, you need to be alone in front of the webcam and stay there the entire time the video is being played.

**Multiple-Faces**: It appears that you are not watching the content alone. Please make sure that you alone are in front of the webcam and start watching the video again.

**Numerous-No-face**: We could not find you in front of the webcam. Please rewatch the video while staying in front of the webcam.

**Tired+Unfocused**: It appears that you have not chosen the right time to watch the videos and may lack sufficient concentration. Please drink coffee/tea and then rewatch the video.

**Tired+Confused**: It appears that you are a little tired and find the content confusing. Please take a rest or drink coffee/tea before continuing.

**Unfocused+Confused**: (1) It appears that you find the content a little confusing. This could be because you have not focused enough on watching the video. You might find it helpful to find a quiet place and put away your mobile phone before rewatching the video. (2) It appears that you find the content a little confusing. This could be because you have not focused enough on watching the video. You might find it helpful to find a quiet place and put away your mobile phone before rewatching the video. You are also recommended to watch the following supplementary videos to better understand the subject.

**Engaged+Tired**: Thank you for paying attention to the videos, but you seem to be a little tired. Please take a rest before proceeding to the next video.

**Engaged+Confused**: (1) Thank you for paying attention to the videos, but it appears that you find some of the content confusing. Please watch the video again. (2) Thank you for paying attention to the videos, but it appears that you need further explanation about some of the content. You can find these explanations and more examples in the following supplementary videos.

**Disengaged+Confused**: (1) It appears that you have trouble understanding some parts of the lesson. Please watch the video again. (2) It appears that you have trouble understanding some parts of the lesson. You can find more examples in the following supplement videos.

**Tired+No-Face**: It appears that you are a little tired and cannot stay in front of the webcam. If you feel tired, please take a rest before continuing or return another time.

**Tired+Disengaged**: If you feel that the content is not right for you, it could be because you are tired. Please take a rest and return another time.

**Engaged+No-Face**: Thank you for paying attention to the content, but it appears you cannot stay in front of the webcam. Watching all parts of the video will have a great impact on your learning.

**Disengaged+No-Face**: If you do not want to watch videos now or cannot stay in front of the webcam, it is recommended that you return another time.

**Engaged+Unfocused**: We are glad you enjoy watching the video, but also suggest that you should watch the videos more carefully.

**No-Face**: It appears that you have not stayed in front of the webcam during the lesson. For better learning, please remain in front of the webcam while watching the video.

**Unfocused**: It appears you are not focused enough while watching the video. Putting aside your cell phone and being in a quiet environment may help you remain focused on the lesson.

**Tired**: You appear to be a little tired. Please take a rest before continuing or return another time.

**Engaged**: Excellent! Keep it up.

**Confused**: (1) It appears that you find some of the content confusing. Please rewatch the video more carefully to fully understand the contents. (2) It appears that you find some of the content confusing. Please check the additional explanations and examples provided in the following supplementary videos.

**Disengaged**: If you feel this is not the right time for you to engage in this learning session, you may return another time.

**Neutral**: You can proceed to the next video.

## D. Questions of the final tests

### D.1. First session

- What kind of artificial intelligence is designed to do specific tasks like spam email detection, speech recognition, etc.?

  1. Artificial narrow intelligence
  2. Artificial general intelligence

- What is the name of the artificial intelligence technology that maps inputs to outputs?

  1. Artificial general intelligence
  2. Reinforcement learning
  3. Supervised learning
  4. Unsupervised learning

- Imagine that you are building a system of speech recognition based on supervised learning. Which of the following would be the best choice for maximizing performance?

  1. A medium-sized neural network with large datasets
  2. A small-sized neural network with small datasets
  3. A medium-sized neural network with small datasets
  4. A large-sized neural network with large datasets

- Which of the following statements is true in regard to supervised learning?

  1. The only way to collect data for supervised learning is manual tagging.
  2. Only structured data (e.g. tables) are of value for supervised learning.
  3. It does not matter how we collect data. The more data, the better.
  4. Some data are more valuable than others. Using an artificial intelligence team can help us collect these data.

- Which of the following would be a problem when working with data?

  1. Missing parts
  2. Wrong labels
  3. Wrong data type (structured/unstructured)
  4. All of the above

- Imagine that you are working in a motorcycle manufacturing company. Which of the following would be part of the unstructured data in your company?

  1. Audio files of engine noises - Maximum speeds of motorcycles
  2. Images of motorcycles - Audio files of engine noises
  3. Maximum speeds of motorcycles - Weekly sales records of the past year
  4. Images of motorcycles - Weekly sales records of the past year

**D.2. Second session**

- Which of the following can be a data science project for launching a website selling dog food?

  1. Collecting a large number of pictures of dogs and dog-related stuff.
  2. Building a neural network that accurately mimics how a dog's brain works
  3. Creating a slide showing how sales can be improved by changing prices
  4. Building a system that detects dog food in pictures

- Which of the following is incorrect?

  1. Artificial intelligence is a type of deep learning.
  2. The terms "neural network" and "deep learning" refer to the same thing and are used interchangeably.
  3. Machine learning is a subfield of artificial intelligence.
  4. Despite being similar in some respects, artificial intelligence and data science are two different things.

- What should an artificial intelligence company be proficient in?
  1. Data collection strategies
  2. Building integrated databases
  3. Identification of task automation opportunities
  4. All of the above

- Which of the following can now be easily done by artificial intelligence?

  1. Recognizing the emotional state of people from their face in pictures
  2. Building a system capable of collecting all economic news of the month and preparing a fifty-page stock market forecast report
  3. Building a personal assistant capable of empathizing with the user after examining their conversations
  4. Building a system capable of proving scientific theorems